\documentclass[sigconf]{acmart}

\renewcommand\footnotetextcopyrightpermission[1]{} 
\usepackage{balance}
\usepackage{bm}

\def\BibTeX{{\rm B\kern-.05em{\sc i\kern-.025em b}\kern-.08emT\kern-.1667em\lower.7ex\hbox{E}\kern-.125emX}}
    
\copyrightyear{2020} 
\acmYear{2020} 
\setcopyright{acmcopyright}
\acmConference[WSDM '20]{The Thirteenth ACM International Conference on Web Search and Data Mining}{February 3--7, 2020}{Houston, TX, USA}
\acmBooktitle{The Thirteenth ACM International Conference on Web Search and Data Mining (WSDM '20), February 3--7, 2020, Houston, TX, USA}
\acmPrice{15.00}
\acmDOI{10.1145/3336191.3371834}
\acmISBN{978-1-4503-6822-3/20/02}

\begin{document}

\fancyhead{}

\title{Popularity Prediction on Social Platforms with Coupled Graph Neural Networks}

\author{Qi Cao$^{1,2}$, Huawei Shen$^{1,2}$, Jinhua Gao$^1$, Bingzheng Wei$^3$, Xueqi Cheng$^1$}
\email{{caoqi, shenhuawei, gaojinhua, cxq}@ict.ac.cn,     coltonwei@tencent.com}
\affiliation{
  \institution{ $^1$ CAS Key Laboratory of Network Data Science and Technology, Institute of Computing Technology,}
  \city{Chinese Academy of Sciences, Beijing, China}
}
\affiliation{
  \institution{ $^2$ University of Chinese Academy of Sciences, Beijing, China}
}
\affiliation{
  \institution{ $^3$ WeChat, Tencent Inc}
}

%
\renewcommand{\shortauthors}{Qi Cao, et al.}

%
\begin{abstract}
Predicting the popularity of online content on social platforms is an important task for both researchers and practitioners. Previous methods mainly leverage demographics, temporal and structural patterns of early adopters for popularity prediction. However,
most existing methods are less effective to precisely capture the \emph{cascading effect} in information diffusion, in which early adopters try to activate potential users along the underlying network. In this paper, we consider the problem of \emph{network-aware popularity prediction}, leveraging both early adopters and social networks for popularity prediction. We propose to capture the cascading effect explicitly, modeling the activation state of a target user given the activation state and influence of his/her neighbors.
To achieve this goal, we propose a novel method, namely CoupledGNN, which uses two coupled graph neural networks to capture the interplay between node activation states and the spread of influence.
By stacking graph neural network layers, our proposed method naturally captures the cascading effect along the network in a successive manner. Experiments conducted on both synthetic and real-world Sina Weibo datasets demonstrate that our method significantly outperforms the state-of-the-art methods for popularity prediction.
\end{abstract}

%
%

%

%
\maketitle


\section{Introduction}
With the booming of social media platforms, e.g., Twitter, Facebook, Sina Weibo, Instagram, etc, the production and dissemination of user-generated online content, which we refer to as a piece of \emph{information}, becomes extremely convenient and common in our life. Every day, there are tens of millions of information generated on these platforms~\cite{Cao:2017:DBG:3132847.3132973}. With such a vast amount of information, predicting the \emph{popularity} of pieces of information is valuable for us to discover the hot information in advance and to help people out of the dilemma of information explosion. However, due to the openness of social platforms and the cascading effect of information diffusion, it's very challenging to accurately predict the popularity of online content.

\begin{figure}
  \centering
  \includegraphics[width=3.75in]{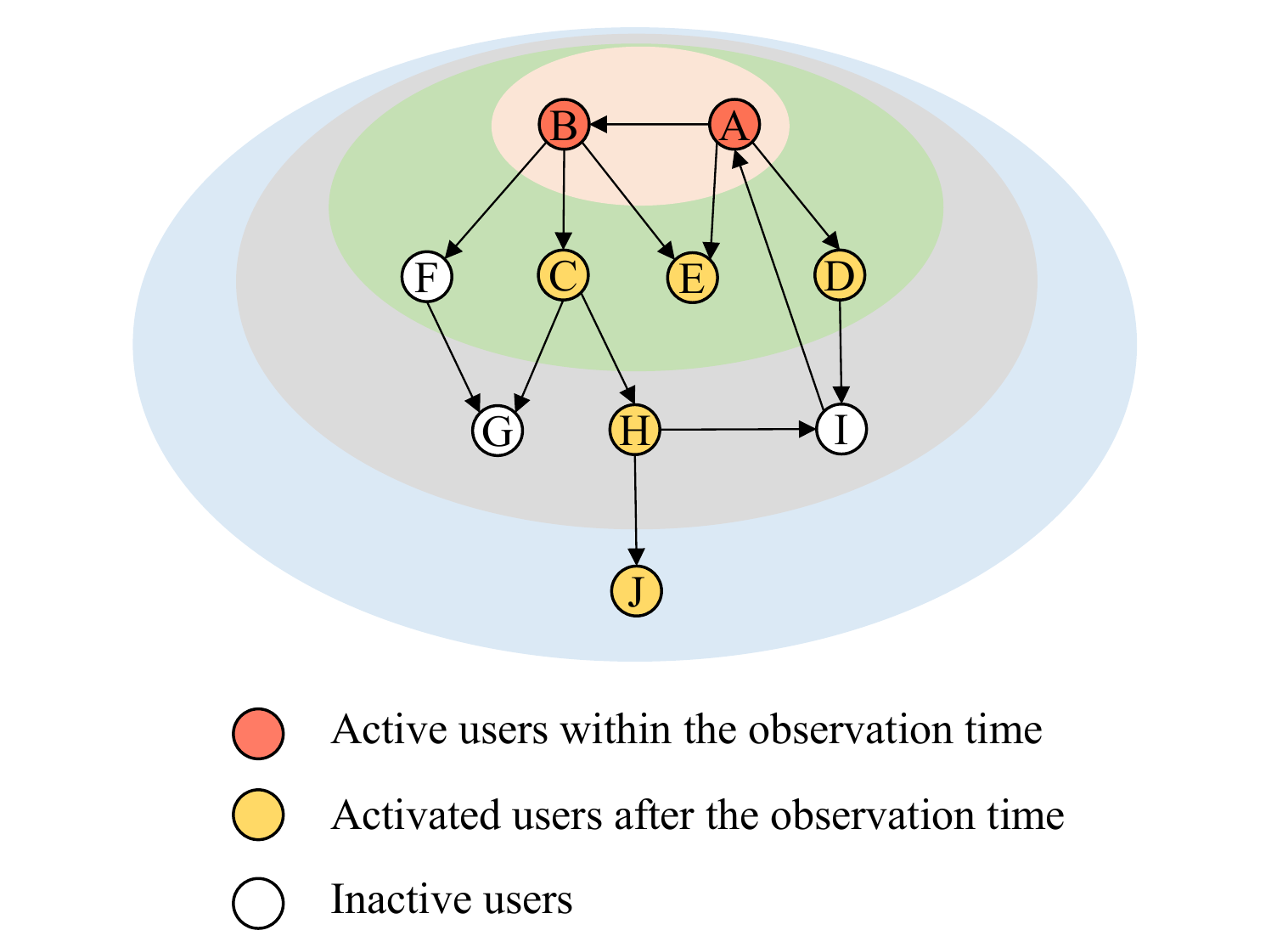}\\
  \caption{\label{motivation}Example of cascading effect in information diffusion. The light orange circle represents the subgraph of early adopters. The green, gray, and blue circles represent the neighborhoods of early adopters that are reachable within one-hop, two-hops, and three-hops respectively.} 
\end{figure}

In the past decade, a series of efforts have been devoted to the popularity prediction problem in social networks, considering this problem either as a regression ~\cite{Zhao2015seismic,Cao:2017:PPO:3041021.3054242,Li2017deepcas} or classification task~\cite{weng2014predicting,WeiXin2019AAAI}. Generally speaking, popularity prediction aims to predict future popularity when observing early adopters at a specific observation time (see the light orange circle in Figure~\ref{motivation} as an example). Various hand-crafted features of early adopters are extracted to predict the future popularity, e.g., demographic features in user profile, user activity~\cite{Gao:2019:TEM:3319626.3301303}, user degree~\cite{Lerman:2008:degree}, density of the subgraph of early adopters~\cite{guille2012predictive,Xiao2014edgedensit}, as well as substructure~\cite{Ugander5962substructures} and community~\cite{weng2014predicting,PhysRevE.70.066111community}. With the success of representation learning methods, end-to-end deep representation learning methods are also proposed to automatically learn the representation of the subgraph of early adopters~\cite{Li2017deepcas}. In addition, to further improve the prediction performance, temporal information~\cite{Pinto:2013:UEV:2433396.2433443,Shen2014RPP,wang2017topological} of the early adopters and the content information~\cite{ZhangWWW2018images,WeiXin2019AAAI} are further utilized. The methods mentioned so far mainly focus on the characteristics of early adopters or the subgraph of early adopters, ignoring the cascading effect (shown in Figure 1) in information diffusions which is one of the keys to accurately predict future popularity of online content in social platforms. 

To further characterize the cascading effect, researchers have also made some attempts. They adopt some statistics, such as the average number of fans of users, to approximate the impact of cascading effect in each generation~\cite{Zhao2015seismic,Mishra:2016:FDP:2983323.2983812,yu2015micro}. However, since they only adopt simple statistics and regardless of the explicit network structure governing the cascading effect, they are less effective for popularity prediction.

In this paper, we focus on the \emph{network-aware popularity prediction} problem, leveraging both early adopters and network structure for predicting the popularity of online content on social platforms. 
To effectively capture the crucial cascading effect, we devote to applying graph neural networks to successively characterize the activation state of each user. Specifically, the activation of a target user is intrinsically governed by two key components, i.e., the state of neighbors and the spread of influence, along social networks. 
As a result, we propose to model the iterative interplay between node states and the spread of influence by two coupled graph neural networks. One graph neural network models the spread of interpersonal influence, gated by the activation state of users. The other graph neural network models the update of the activation state of each user via interpersonal influence from their neighbors. With the iterative aggregation mechanism of the neighborhood in graph neural networks, the cascading effect along the network structure is naturally characterized. Note that, other information like temporal or content, if available, can be further included in the prediction model by representation fusion flexibly. 

We verify the effectiveness of our proposed coupled graph neural networks on both the synthetic data and real-world data in Sina Weibo. Experimental results demonstrate that our proposed method significantly outperforms all the state-of-the-art methods. For conveience of the reproduction of the results, we have made the source code publicly available\footnote{https://github.com/CaoQi92/CoupledGNN.}.

\section{Related Work}
In this section, we briefly review the research on the popularity prediction, traditional diffusion models, the development and application of graph neural networks.

\subsection{Popularity Prediction}
Popularity prediction aims to predict the future popularity of online content when observing early adopters within the observation time. Due to the openness of social platforms and the cascade phenomenon of online content, future popularity results in huge variance and is challenging to predict. The predictability of particular types of information has been proved to some extent, e.g., tweet/microblogs~\cite{Martin2016twitter,Gao:2019:TEM:3319626.3301303}, images~\cite{ZhangWWW2018images}, videos\cite{RizoiuWWW2017video}, recipes~\cite{Sanjocikm2017recipe}, and academic papers~\cite{Shen2014RPP}. 

Generally speaking, existing methods for popularity prediction mainly focus on four types of information, i.e., content, temporal information, early adopters and network structure. For content information, hierarchical attention networks~\cite{WeiXin2019AAAI} or user-guided hierarchical attention mechanisms~\cite{ZhangWWW2018images} are proposed to characterize the content features. For temporal information, heuristical temporal features~\cite{Pinto:2013:UEV:2433396.2433443}, time series models including recurrent neural network~\cite{Wu:2018:ATM:3269206.3271714} and temporal convolutional network~\cite{shao2019temporal}, or point process method including reinforced Poisson processes~\cite{Shen2014RPP} and Hawkes process~\cite{Zhao2015seismic,Mishra:2016:FDP:2983323.2983812,RizoiuWWW2017video,Yan:2018:IML:3304889.3305070}, are proposed to devote to capture the underlying laws or patterns behind the temporal information.

As for early adopters and network structure, which is also the focus of this paper, both feature-based methods and representation learning methods are proposed. The designed effective features in the former one including node degree~\cite{Lerman:2008:degree,Zhao2015seismic}, the number of nodes in the frontier graph~\cite{Guo:2016:CMC:frontiers}, cascade density~\cite{guille2012predictive,Xiao2014edgedensit}, substructures~\cite{Ugander5962substructures}, community~\cite{weng2014predicting,PhysRevE.70.066111community} 
and so on. Unfortunately, the performance of such methods heavily depends on the quality of the hand-crafted features, which are generally extracted heuristically. To avoid the above heuristic feature extraction process, attempts of end-to-end deep representation learning fashion are proposed to automatically learn the impact representation of early adopters by cascading effect~\cite{Li2017deepcas,wang2017topological,Cao:2017:DBG:3132847.3132973}. 

However, the above methods are less effective to capture the cascading effect in information diffusion, since they neglect the explicit interactions between users along the underlying social network. In contrast, the method proposed in this paper effectively capture such cascading effect along the network structure by coupled graph neural networks.

\subsection{Diffusion Models}
Modeling how information diffuse is of outstanding interest over the past decades. There are two classic diffusion models in this category, i.e., Independent Cascades (IC) model~\cite{goldenberg2001talk} and Linear Threshold (LT) model~\cite{Granovetter:10.1086/226707}. The diffusion process of these models is both iteratively carried on a synchronous way along discrete time steps from initial adopters. The synchronicity assumption is further relaxed by proposing asynchronous continuous-time extensions~\cite{guille2012predictive,Gomez-Rodriguez:2011:UTD:3104482.3104553}. Such diffusion models can well capture the cascading effect along network by iteratively modeling the specific activation process. However, they generally need an extremely high number of Monte-Carlo simulations to estimate the final influence spread, i.e., the popularity to be predicted. Such a prediction process is time-consuming and limits its applicability to real scenarios. 

The difference between our proposed method and works of this line is that we do not model the specific diffusion process, but utilize the graph neural networks to directly model the influence of cascading effect by neighborhood aggregation, which is more efficient and flexible.

\begin{figure*}
  \centering
  \includegraphics[width=7.15in]{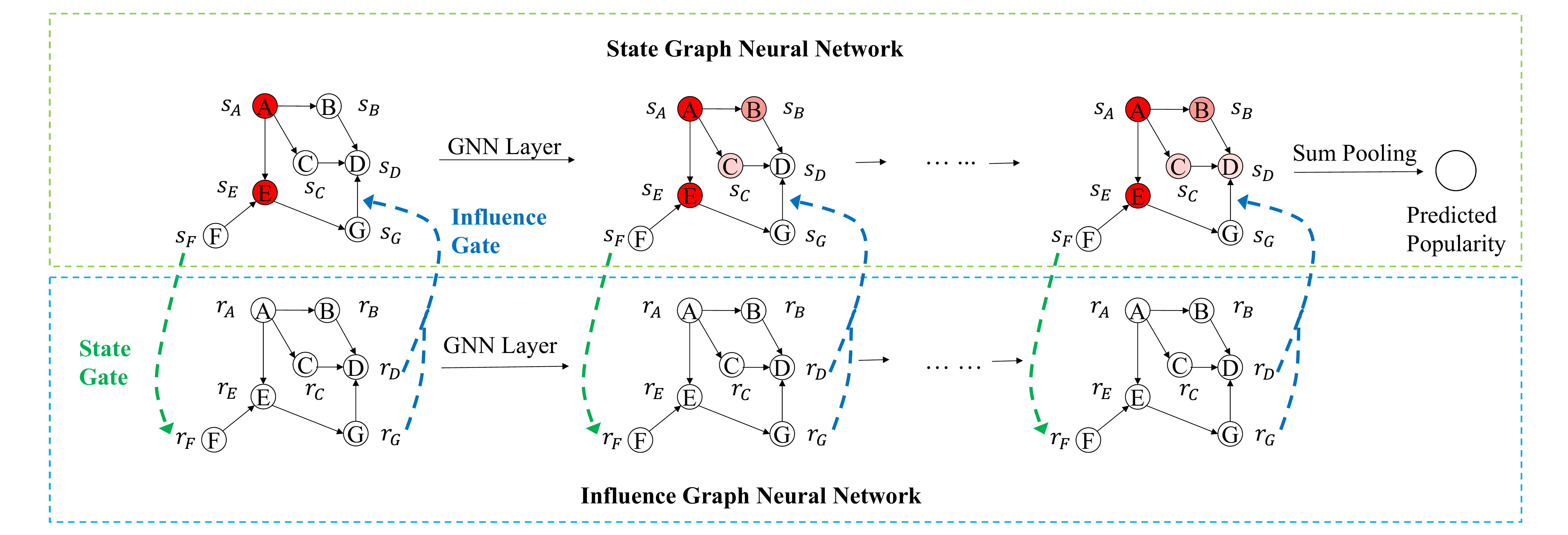}
  \caption{\label{framework}The framework of coupled graph neural networks for popularity prediction. $s_*$ and $r_*$ are the activation state and influence representation of user $*$ respectively.}
  \Description{The 1907 Franklin Model D roadster.}
\end{figure*}

\subsection{Graph Neural Networks}
Inspired by the huge success of neural networks in Euclidean space, recently there has been a surge of interest in graph neural network approaches for representation learning of graphs~\cite{NIPS2017_6703,velivckovic2018graph,bingbing2019graphwave,xu2019powerful}. Graph neural networks (GNNs) broadly follow a recursive neighborhood aggregation fashion, where each node updates its representation by aggregating the representation of its neighborhood. After $K$ iterations of aggregation, the updated representation of each node captures both the structural and representation information within the node's $K$-hop neighborhood~\cite{xu2019powerful}.

GNNs have been successfully applied to a lot of non-Euclidean domain problems, e.g., semi-supervised learning on graph~\cite{kipf2017semi,Xu:2019:GCN:3367243.3367306}, social influence prediction~\cite{Qiu2018deepinf}, correlated temporal sequence modeling~\cite{shang2019geometric}. Among the above, the application to social influence prediction, i.e., DeepInf~\cite{Qiu2018deepinf}, is the most related one with our work. However, since DeepInf more focuses on the prediction of the micro action status of a user on a fixed-sized local network, rather than the macro popularity prediction on the global diffusion network studied in this paper, it performs not well for future popularity. 

In this paper, we devote to utilizing GNNs to characterize the cascading effect in popularity prediction. To better adapt to the scenario of information diffusion, we design a novel model coupled graph neural networks to solve the popularity prediction problem.

\section{Preliminaries}
This section gives the formal definition of the popularity prediction problem studied in this paper and the general framework of GNNs.

\subsection{Problem Definition}
Supposing that we have $M$ pieces of information, the observed cascade of information $m$ is recorded as the set of early adopters within the observation time window $T$, i.e., $\mathcal{C}_T^m=\{u_1,u_2,...,u_{n_T^m}\}$, where $n_T^m$ is the total number of adopted or active users of information $m$ within the observation time window $T$. For example, the observed cascade in Figure~\ref{motivation} is recorded as $\mathcal{C}_T^m = \{A,B\}$. In addition to the observed cascades, given the underlying network which governing the information diffusion, e.g., the following relationships in Sina Weibo, we can formalize the popularity prediction problem studied in this paper as:

\textbf{Network-aware Popularity Prediction.} Given the observed cascades $\mathcal{C}_T^m$ and the underlying network $\mathcal{G} = (\mathcal{V},\mathcal{E})$, where $\mathcal{V}$ is the set of all users, $\mathcal{E}\subseteq \mathcal{V}\times \mathcal{V}$ is the set of all relationships between users, this problem aims to predict the final popularity of information $m$, i.e., $n_\infty ^m$.

Note that, the network-aware popularity prediction problem emphasizes the role of the network, i.e., there are interactions between early adopters and potential active users, or among the potential active users. It's precisely because of this characteristic, making the capture of the cascading effect along the network becomes the key to accurately predict the future popularity of online content.

\subsection{General Framework of GNNs}
GNNs is an effective framework for representation learning of graphs. As introduced in Section 2.3, many variants of graph neural networks have been proposed. They usually follow a neighborhood aggregation strategy, where the representation of a node is updated by recursively aggregating the representation of its neighboring nodes. Formally, the $k-$th layer of a graph neural network is generally formulated as in~\cite{xu2019powerful}:
\begin{equation}
    \bm{a}_v^{(k)} = \mbox{AGGREGATE}\left (\left \{\bm{h}_u^{(k)}:u\in \mathcal{N}(v)\right \}\right ),
\end{equation}
\begin{equation}
    \bm{h}_v^{(k+1)} = \mbox{COMBINE}\left (\bm{h}_v^{(k)},\bm{a}_v^{(k)}\right ),
\end{equation}
where $\bm{h}_v^{(k)}$ is the feature vector of node $v$ at the $k$-th layer, $\mathcal{N}(v)$ is the set of nodes which appear in the neighborhood of node $v$. The choice of the function $\mbox{AGGREGATE}(*)$ and $\mbox{COMBINE}(*)$ in GNNs is crucial.

The representation of the entire graph $\bm{h}_{\mathcal{G}}$ is obtained by a READOUT function:
\begin{equation}
    \bm{h}_{\mathcal{G}} = \mbox{READOUT}\left (\left \{\bm{h}_v^{(k+1)}:v\in \mathcal{G} \right \}\right ).
\end{equation}

$\mbox{READOUT}(*)$ can be a simple function such as summation or a more sophisticated graph-level pooling function~\cite{NIPS2018_7729,zhang2018end}.

\section{Methods}
In this section, we introduce the proposed coupled graph neural network (CoupledGNN) for network-aware popularity prediction. We design the CoupledGNN model to capture the \emph{cascading effect} widely observed in information diffusion over social networks. 

\subsection{Framework of CoupledGNN}
The cascading effect indicates that the activation of one user will trigger its neighbors in a successive manner, forming an information cascade over social networks. For a target user, whether he/she could be activated is intrinsically governed by two key components, i.e., \emph{the state of neighbors} and \emph{the spread of influence}, direct or indirect, over social networks. In this sense, the cascading effect is intrinsically the iterative interplay between node states and the spread of influence. Previous methods, e.g., independent cascade model, assumes a fixed-yet-unknown interpersonal influence and probes the interplay manifested as the cascading effect over the social network via Monte-Carlo simulation. 
 
In this paper, we propose to use two coupled graph neural networks to naturally capture the cascading effect, or more specifically, the interplay of node states and the spread of influence. One graph neural network, namely state graph neural network, is used to model the activation state of nodes. The other graph neural network, namely influence graph neural network, is used to model the spread of influence over social networks. The two graph neural networks are coupled through two gating mechanisms. The framework of our coupled graph neural networks is shown in Figure~\ref{framework}.

\subsection{State Graph Neural Network}
\begin{figure}
  \centering
  \includegraphics[width=2.2in]{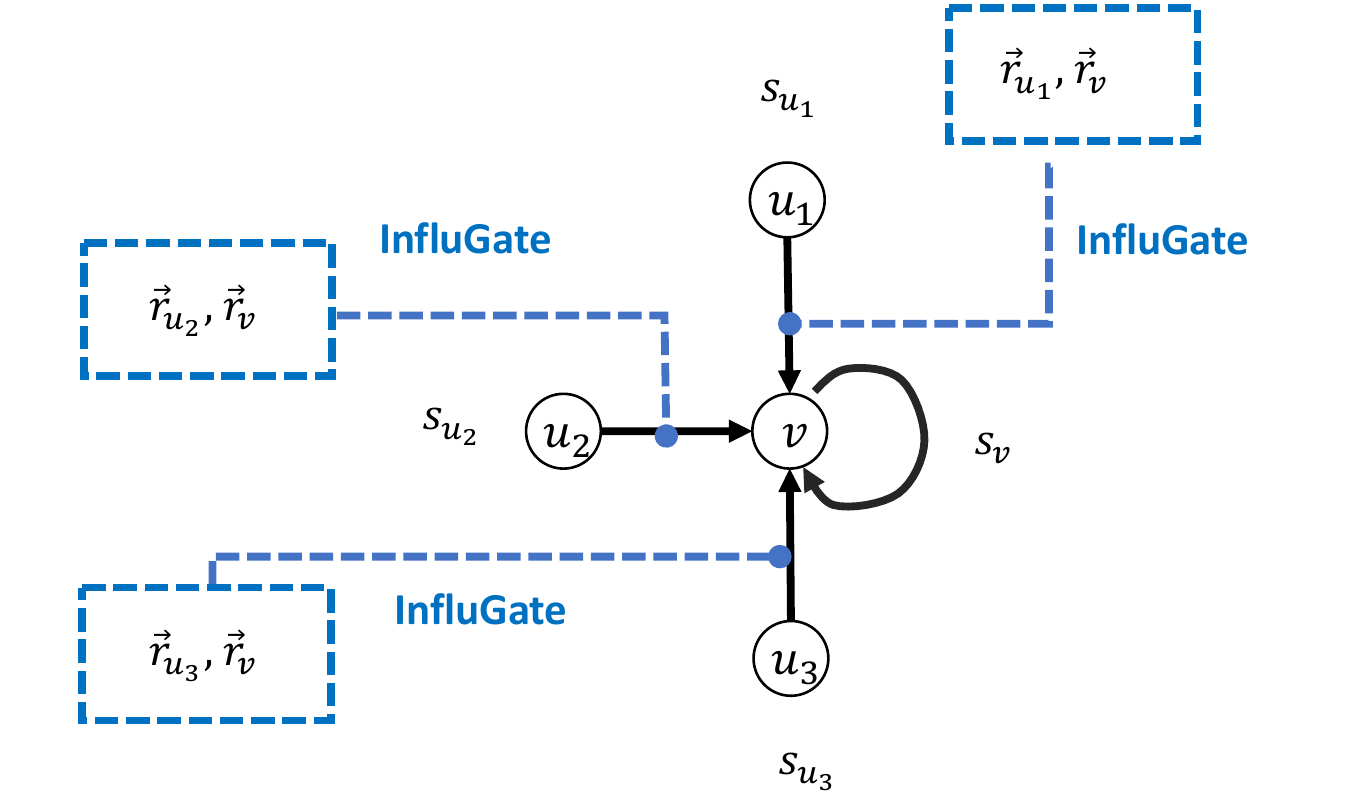}
  \caption{\label{state-gnn}Mechanisms of state graph neural network.}
\end{figure}

The state graph neural network is to model the activation of each user during the cascading effect. Specifically, for a target user $v \notin \mathcal{C}_T^m$, since he/she is usually influenced by the active users in the neighborhood $\mathcal{N}(v)$, we apply a graph neural network to model the activation state of each user (shown in Figure~\ref{state-gnn}). Each user is associated with a one-dimensional value $s_v$, indicating the activation state of user $v$. Besides, since the interpersonal influence between the pair of users is generally various, we model such heterogeneous influence weight by an influence gating mechanism, i.e., 
\begin{equation}
\mbox{InfluGate}\left (\bm{r}_u^{(k)} , \bm{r}_v^{(k)}\right ) = \bm{\beta}^{(k)} [\bm{W}^{(k)}\bm{r}_u^{(k)} \parallel \bm{W}^{(k)}\bm{r}_v^{(k)}],
\end{equation}
where $\bm{r}_u^{(k)} \in \mathbb{R}^{h^{(k)}} $ is the influence representation of user $u $ at the $k$-th layer, $\bm{W}^{(k)}\in \mathbb{R} ^{h^{(k+1)}\times h^{(k)}}$ is a weight matrix to transform the influence representation from dimension $h^{(k)}$ to $h^{(k+1)}$, $\bm{\beta}^{(k)} \in \mathbb{R}^{2h^{(k+1)}}$ is a weight vector. Note that, Equation $4$ is just one instance of the $\mbox{InfluGate}(*)$ function. We can also choose other types of functions which can reflect the influence gating between the pair of users.

After obtaining the heterogeneous influence weight by influence gating function, the aggregation of the expected influence that the target user $v$ receives from his/her neighborhood is:

\begin{equation}
  a_v^{(k)} = \sum_{u\in 
  \mathcal{N}(v)} \mbox{InfluGate}\left (\bm{r}_u^{(k)},\bm{r}_v^{(k)}\right )s_u^{(k)}+p_v,
\end{equation}
where $\mathcal{N}(v)$ is the neighborhood of user $v$, $\mbox{InfluGate}\left (*\right )s_u^{(k)}$ is the expected influence considering the activation state $s_u^{(k)}$ of neighbor $u$, $p_v \in \mathbb{R}$ is a self activation parameter to reflect the probability that user $v$ may be activated by ways out of following relationships, e.g., offline communication or browsing the hot list in the front pages. Note that, Equation 5 is actually a specific design of the AGGREGATE function as mentioned in Section 3.2.

Then the $\mbox{COMBINE}(*)$ function used to update the activation state of user $v$ is defined as the weighted sum of the neighborhood aggregation and the activation state of user $v$ itself:
\begin{equation}
s_v^{(k+1)} = \left\{
\begin{array}{ll}
    1, &v\in \mathcal{C}_T^m  \\
    \sigma \left( \mu_s^{(k)} s_v^{(k)}+\mu_a^{(k)} a_v^{(k)} \right ), & v \notin \mathcal{C}_T^m
\end{array},
\right.
\end{equation}
where $\mu_s^{(k)}, \mu_a^{(k)} \in \mathbb{R}$ are weight parameters, $\sigma$ is a nonlinear activation function. The initial activation state of user $v$ is defined as 
\begin{equation}
s_v^{(0)} = \left\{
\begin{array}{ll}
    1, &v\in \mathcal{C}_T^m  \\
    0, & v \notin \mathcal{C}_T^m
\end{array}.
\right.
\end{equation}

\subsection{Influence Graph Neural Network}
\begin{figure}
  \centering
  \includegraphics[width=2.35in]{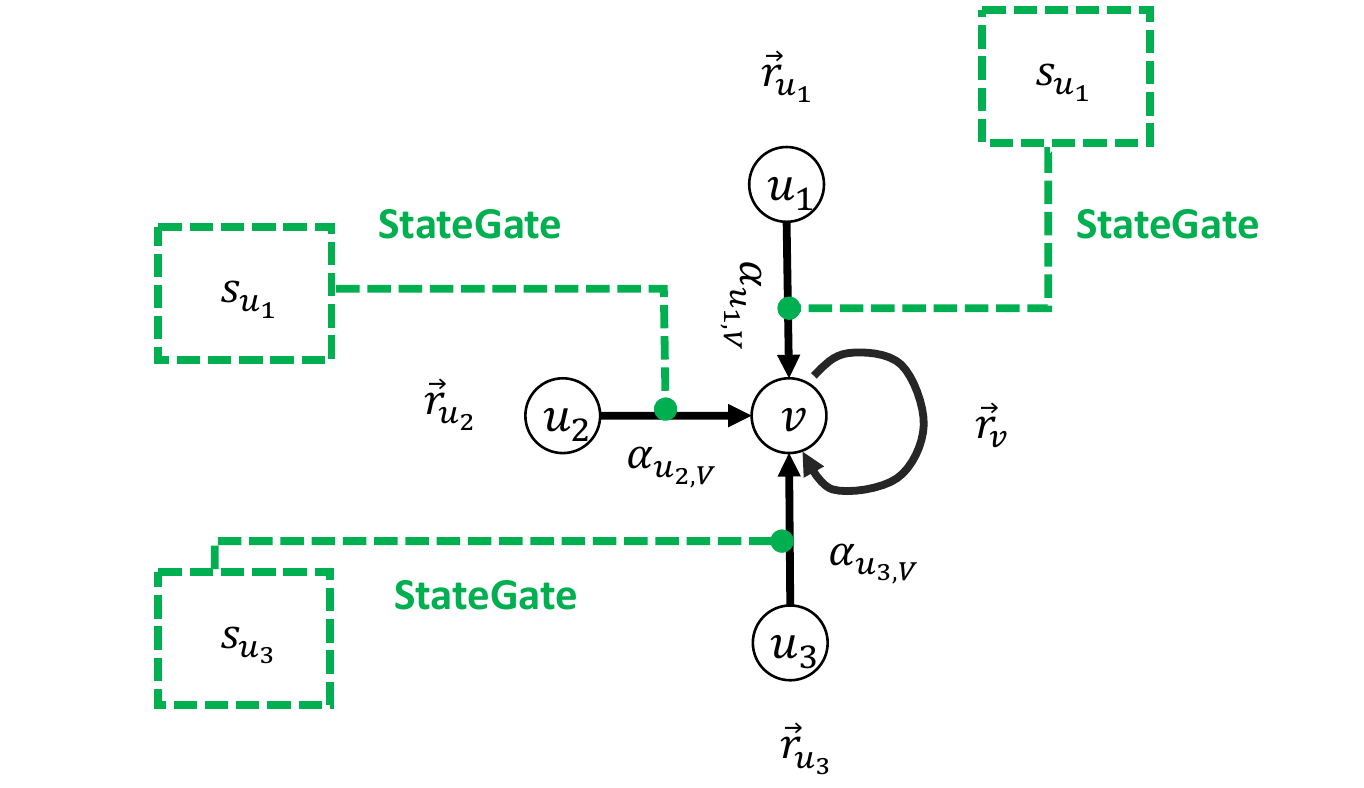}
  \caption{\label{influence-gnn}Mechanisms of influence graph neural network.}
\end{figure}

The influence graph neural network is to model the diffusion of interpersonal influence in the social network. Specifically, each user $v$ is associated with an influence representation $\bm{r}_v$. Then the influence representation of active users further diffuses to other users along with network structure, implemented by neighborhood aggregation of graph neural networks and a state gating mechanism.
The entire mechanism of influence graph neural network is shown in Figure~\ref{influence-gnn}. Specifically, the neighborhood aggregation is defined as:
\begin{equation}
  \bm{b}_v^{(k)} = \sum_{u\in 
  \mathcal{N}(v)}  \mbox{StateGate}\left (s_u^{(k)}\right ) \alpha_{uv}^{(k)} \bm{W}^{(k)} \bm{r}_u^{(k)},
\end{equation}
where $\bm{r}_u^{(k)}\in \mathbb{R}^{h^{(k)}}$ is the influence representation of user $u$ at $(k)$-th layer, $\mathbf{W}^{(k)}\in \mathbb{R} ^{h^{(k+1)}\times h^{(k)}}$ is a weight matrix to transform the influence representation from dimension $h^{(k)}$ to $h^{(k+1)}$. $\mbox{StateGate}(*)$ is the state gating mechanism, implemented by a 3-layer MLP in this paper to reflect the nonlinear effect of state. $\alpha_{uv}^{(k)}$ is the attention weight from user $u$ to user $v$, where we adopt the formulation used in \cite{velivckovic2018graph}, i.e., 
\begin{equation}
e_{uv}^{(k)}=\bm{\gamma}^{(k)} [ \bm{W}^{(k)}\bm{r_u}^{(k)}\parallel \bm{W}^{(k)}\bm{r}_v^{(k)} ], 
\end{equation}
\begin{equation}
\alpha_{uv}^{(k)} = \mbox{softmax}(e_{uv}^{(k)}) = \frac{\mbox{exp}(e_{uv}^{(k)})}{\sum_{z\in \mathcal{N}(v)}\mbox{exp}(e_{zv}^{(k)})},
\end{equation}
where $\bm{\gamma}^{(k)} \in \mathbb{R}^{2h^{(k)}}$ is a weight vector.

Then the influence representation of user $v$ at $(k+1)$-th layer is updated by
\begin{equation}
  \bm{r}_v^{(k+1)} = \sigma \left ( \zeta_r^{(k)} \bm{W}^{(k)}\bm{r}_v^{(k)}+\zeta_b^{(k)} \bm{b}_v^{(k)} \right ),
\end{equation}
where $\zeta_r^{(k)}, \zeta_b^{(k)} \in \mathbb{R}$ are weight parameters, $\sigma$ is a nonlinear activation function. The initial influence representation $\bm{r}_v^{(0)}$of user $v$ used in this paper consists of two parts: node embeddings and node features. We will discuss each part in detail in the section of implementation details.

\subsection{Output Layer}
After $K$ layers of graph neural networks for both activation state and influence representation, the output activation probability of each user in the network is $s_v^{(K)} \in [0,1]$, i.e., the output of the last layer in the state graph neural network.
The popularity to be predicted is then obtained by a sum pooling mechanism over all users in the network, i.e., 
\begin{equation}
  \hat{n}^m_\infty = \sum_{u\in \mathcal{V}} s_u^{(K)}
\end{equation}

As for the loss function to be optimized, we consider the mean relative square error (MRSE) loss~\cite{Tatar2014survey,Cao:2017:PPO:3041021.3054242}, which is robust to outliers as well as smooth and differentiable:
\begin{equation}
  L_{\mbox{MRSE}} = \frac{1}{M} \sum_{m=1}^{M}\left (\frac{\hat{n}_\infty ^m-n_\infty ^m}{n_\infty ^m}\right )^2,
\end{equation}
where $M$ is the total number pieces of information, $n_\infty ^m $ is the true final popularity of information $m$.

To avoid over-fitting and accelerate the process of convergence, we also add a L2 and user-level cross entropy to the objective function as regularization: 

\begin{equation}
  L = L_{\mbox{MRSE}} +L_{\mbox{Reg}},
\end{equation}
where 
\begin{math}
  L_{\mbox{Reg}} =\eta \sum_{p \in \mathcal{P}}\parallel p \parallel _2 + \lambda L_{\mbox{user}},
\end{math}
$ \mathcal{P}$ is the set of parameters, $\eta$ and $\lambda$ are hyper-parameters. $L_{user}$ is the user-level cross entropy, i.e., 
\begin{math}
L_{\mbox{user}} = \frac{1}{M} \sum_{m=1}^{M}\frac{1}{|\mathcal{V}|} \sum_{v\in \mathcal{V}}  \left (s_v^{\infty}\log s_v^{(K)} + (1-s_v^{\infty })\log s_v^{(K)}\right ) ,
\end{math}
$s_v^{\infty}$ is the true final activation state of each user.

\subsection{Computational Complexity}
For the state graph neural network, the computational complexity including the influence gating mechanism at $k$-th layer, i.e., $\mathcal{O}(|\mathcal{V}|h^{(k-1)}h^{(k)}+|\mathcal{E}|h^{(k)})$, and  the updation of activation state, i.e., $\mathcal{O}(\mathcal{|V|}+\mathcal{|E|})$. For the influence graph neural network, the computational complexity at $k$-th layer is $\mathcal{O}(|\mathcal{V}|+|\mathcal{V}|h^{(k-1)}h^{(k)}+|\mathcal{E}|h^{(k)})$. Sum up, the computational complexity of coupled graph neural network is $\mathcal{O}(p|\mathcal{V}|+q|\mathcal{E}|)$, where $p,q$ are small constant associated with the hidden dimension $h^{(k)}$ at each layer. It's worth noting that the above computational complexity is based on the computation of the whole network. To make it more efficient, we can also address several mini-batch with $R$ samples, which makes the algorithm independent of the graph size and achieve $\mathcal{O}(R)$ complexity~\cite{shang2019geometric}.

\section{Experimental Setup}
We compare our CoupledGNN with several state-of-the-art methods on different data sets under various evaluation metrics. The detailed experimental settings are introduced in this section.
\subsection{Data Sets}
To thoroughly evaluate the performance of our methods, we conduct experiments on both the synthetic data set and a real-world data set from Sina Weibo.
\subsubsection{Synthetic Data Set}
The synthetic network is constructed by Kronecker generator~\cite{Leskovec:2010:KGA:1756006.1756039}, which can generate networks that have common structural properties of real networks, i.e., heavy tails for both in- and out-degree distributions, small diameters. The parameter matrix is set to be $[0.9;0.5;0.5;0.1]$ and we retain the largest connected component as the final network, containing $1,086$ nodes and $4,038$ edges. 

As for information cascades, we first sample the seed set of each cascade. The size of the seed set is sampled according to the power-law distribution with parameter 2.5, i.e., $p(n)\propto n^{-2.5}$~\cite{Du:2014:IFL:3044805.3045117}, and the node in each seed set is uniformly sampled. With a given seed set, the commonly used IC model~\cite{goldenberg2001talk} is applied to generate the diffusion data, where the activation probability from node $u$ to node $v$ is set to be $1/d_v$ and $d_v$ is the in-degree of node $v$. 
The observation time window $T$ is set to be 2 time steps in this scenario, i.e., we observe the diffusion process at time step $t=0$ and $t=1$. 

In total, there are $108,600$ information cascades are generated and the cascades with less than 3 active users are filtered out. Finally, $27,218$ information cascades are taken as our data. We randomly sample 80\% of the data as our train set, 10\% as the validation set and 10\% as the test set.

\subsubsection{Sina Weibo Data Set}
For real-world data, let's turn our attention to the information cascades on Sina Weibo, one of the most popular social platform in China. 
The Sina Weibo data set used in this paper is from~\cite{Zhang:2013:SIL:2540128.2540526,Zhang:2015:IYP:2737800.2700398} and publicly available online\footnote{https://www.aminer.cn/Influencelocality.}. The network in this data set is the following network, reflecting the following relationships between users. Note that, such following network is quite related to the retweet information cascades since the posted messages by user B will appear in user A's feed when user A follows user B in Sina Weibo. The following network contains $1.78$ million users and $308$ million following relationships in total. $300$ thousand popular microblog retweet information cascades of these users are included. 
To analysis the retweets behavior of a specific group of users with corresponding the information cascades, we construct a subset of users and messages on the user-microblog bipartite graph. Specifically, we start with a randomly chosen user and then obtain all the messages with coverage $\ge \eta$. The coverage is defined as the number of users in the chosen set normalized by the number of total users in the message. The users appeared in the obtained messages are then added into the chosen user set. We repeated the above steps and obtain $23,732$ users with corresponding $149,53$ information cascades. The largest component of the following network between these users is regarded as the final network, containing $23,681$ users and $1,802,146$ edges. Information cascades with less than 5 active users are filtered out and the remaining $3,228$ pieces of information are taken as our data. As for the observation time window, we set the observation time window $T=1$ hour, 2 hours and 3hours respectively.

\subsection{Baselines}
Since this paper focuses on the network-aware popularity prediction without temporal information, we mainly consider methods that utilize early adopters and network structure as our baselines. 
Existing methods for this problem are mainly classified into two categories: feature-based methods and deep representation learning methods. We choose the state-of-the-art method in each category as our baselines.
Besides, we also include the representative attempt of capturing the cascading effect in Popularity prediction.


\subsubsection{Feature-based.}
We extract all the effective hand-crafted features that can be easily generalized across data sets~\cite{Cheng:2014:CP:2566486.2567997,Gao:2019:TEM:3319626.3301303,Li2017deepcas,shulman2016predictability}. The extracted features are conducted on three types of graphs: the global graph $\mathcal{G}$, the cascade graph $g_c$, and the frontier graph $g_f$. Specifically, the cascade graph $g_c$ contains all early adopters and the corresponding edges between these users. The frontier graph $g_f$ contains all users in the one-hop neighborhood of early adopters and the edges between these neighboring users. As for features, we extract the mean and 90th percentile of the degrees of users~\cite{Lerman:2008:degree}, the number of leaf nodes, edge density~\cite{guille2012predictive,Xiao2014edgedensit} in $g_c$; the number of substructures~\cite{Ugander5962substructures,shulman2016predictability}, including nodes, edges and triangles, and the number of communities and the corresponding coverage of the partition of these communities~\cite{weng2014predicting,PhysRevE.70.066111community} in both $g_c$ and $g_f$. In addition, since the node identity is quite important for popularity prediction~\cite{Li2017deepcas}, here we also include the global node ids in $\mathcal{G}$ as the structure feature. 
Once the cascade is represented as a bag of features, we feed them into a linear regression model with L2 regularization.

\subsubsection{DeepCas~\cite{Li2017deepcas}.} DeepCas is the state-of-the-art deep representation learning method for network-aware popularity prediction, which learns the representation of cascade graphs in an end-to-end manner. 
Specifically, it represents the cascade graph as a collection of sequences by random walks, and then utilizes the embeddings of nodes and recurrent neural networks to obtain the representation of each sequence. Attention mechanisms are further applied to assemble the representation of the cascade graph from sequences. 

\subsubsection{SEISMIC~\cite{Zhao2015seismic}.} SEISMIC is a representative method for attempts of capturing the cascading effect. It is an implementation of Hawkes self-exciting point process and estimates or approximates the impact of cascading effect in each generation by the average number of fans of users.

\subsection{Implementation Details.}
For all baselines and our CoupledGNN model, the hyper-parameters are tuned to obtain the best results on the validation set. The L2-coefficient is chosen from $\{10^{-8}, 10^{-7},...,0.01,0.1\}$.
For feature-based method, since the features of node ids are high-dimensional and sparse, we set a learning rate alone for the parameters of these features and choose from $\Phi_1=\{10^{-5}, 5\times 10^{-5}, 10^{-4},...,0.01\}$. For the parameters of other features, we choose the learning rate from $\Phi_2=\{0.0005,0.001,0.005,0.01\}$. 
Similarly, for DeepCas, the learning rate for user embeddings are chosen from $\Phi_1$, while the learning rate for other parameters are chosen from $\Phi_2$. The user embeddings for DeepCas is initialized by DeepWalk~\cite{deepwalk} which will be further optimized during the training process, while the user embeddings for our CoupledGNN are also obtained by DeepWalk but without further fine-tuning.
The dimension of the embeddings is all set to be 32. The hidden units of RNN in DeepCas is set to be 32, and the units of the first dense layer and the second dense layer in the output part are 32 and 16 respectively. 
As for SEISMIC, we adopt the setting of parameters used in~\cite{Zhao2015seismic}, i.e., setting the constant period $s_0$ to 5 minutes and power-law decay parameters $\theta= 0.242$. Besides, we choose mean degree $n^*$ from $\{1,3,5,10, 20, 50, 100\}$ to minimize the mRSE of validation set. 
For our CoupledGNN model, similar to baselines, the learning rate for self activation parameters of all users are chosen from $\Phi_1$, and the learning rate for other parameters are chosen from $\Phi_2$. The coefficient $\lambda$ in the loss function, which balances the weight of the regularization of user-level cross entropy, is set to be 0.5 in our experiments. The number of GNN layers $K$ is chosen from $\{2,3,4\}$, and each layer contains the same number of hidden units as input. Following~\cite{Qiu2018deepinf}, the vertex features for our CoupledGNN contains coreness, pagerank, hub score, authority score, eigenvector centrality, and clustering coefficient.

\begin{table*}
  \caption{Popularity prediction in Sina Weibo}
  \label{tab:popularity-sina}
  \begin{tabular}{c|c|c|c|c|c|c|c|c|c|c|c|c}
    \toprule
     Observation Time &  \multicolumn{4}{c| }{1 hour}& \multicolumn{4}{c| }{2 hours}& \multicolumn{4}{c }{3 hours}\\
     \midrule
     Evaluation Metric & MRSE &mRSE & MAPE & WroPerc
     & MRSE &mRSE & MAPE  & WroPerc
     & MRSE &mRSE & MAPE  & WroPerc \\
    \midrule
         SEISMIC       &-      &0.2112 &-      &48.63\%
                       &-      &0.1347 &-      &34.59\%
                       &-      &0.0823 &-      &27.15\%\\
         Feature-based &0.2106 &0.1254 &0.3749 &35.17\% 
                       &0.1796 &0.1041 &0.3557 &28.86\%
                       &0.1581 &0.0804 &0.3147 &18.97\% \\
         DeepCas       &0.2077 &\textbf{0.0930} &0.3633 &30.00\%  
                       &0.1650 &0.0670 &0.3134 &20.55\%
                       &0.1365 &0.0361 &0.2813 &17.24\% \\
         CoupledGNN
                       &\textbf{0.1816} &0.0946       
                       &\textbf{0.3515} &\textbf{25.68\%}
                       &\textbf{0.1397} &\textbf{0.0519}       
                       &\textbf{0.2989} &\textbf{17.81\%}
                       &\textbf{0.1120} &\textbf{0.0333}       
                       &\textbf{0.2611} &\textbf{13.01\%}\\

  \bottomrule
\end{tabular}
\end{table*}

\subsection{Evaluation Metrics}
We adopt several different evaluation metrics to comprehensively demonstrate the performance of each method.

\subsubsection{Mean Relative Square Error (MRSE)~\cite{Cao:2017:PPO:3041021.3054242,Tatar2014survey}.} We take the mean relative square error loss also as our evaluation metric for popularity prediction.

\subsubsection{Median Relative Square Error (mRSE)} Since SEISMIC is sensitive to outlier error, we also use median RSE as an evaluation metric, which is defined as the 50th percentile of the distribution of RSE over test data.

\subsubsection{Mean Absolute Percentage Error (MAPE)~\cite{Shen2014RPP,Wu:2018:ATM:3269206.3271714}.} This metric measures the average deviation between the predicted and true popularity. The formulation is 
\begin{equation}
    MAPE= \frac{1}{M}\sum_{m=1}^{M} \frac{|\hat{n}_\infty ^m-n_\infty^m|}{n_\infty^m}.
\end{equation}

\subsubsection{Wrong Percentage Error (WroPerc)} The wrong prediction error is defined as the percentage of online contents that are incorrectly predicted for a given error tolerance $\epsilon$:
\begin{equation}
  WroPerc = \frac{1}{M}\sum_{m=1}^{M}\mathbb{I}[ \frac{|\hat{n}_\infty^m-n_\infty^m|}{n_\infty^m} \ge \epsilon].
\end{equation}
We set the threshold $\epsilon=0.5$ in this paper.

Note that, among all these three evaluation metrics, the smaller the value is, indicating the better the performance of the corresponding method.

\begin{table}
  \caption{Popularity prediction in synthetic dataSet}
  \label{tab:popularity-synthetic}
  \begin{tabular}{c|c|c|c|c}
    \toprule
     Evaluation Metric & MRSE &mRSE & MAPE & WroPerc\\
    \midrule
         SEISMIC       &-      &0.2025 &-      &47.92\%\\
         Feature-based &0.1225 &0.0452 &0.2718 &16.90\% \\
         DeepCas       &0.1199 &0.0361 &0.2657 &16.82\% \\
         CoupledGNN
                    &\textbf{0.1101} &\textbf{0.0339}
                    &\textbf{0.2517} &\textbf{14.71}\%\\

  \bottomrule
\end{tabular}
\end{table}

\section{Experimental Results}
In this section, we first compare our CoupledGNN with baselines on the target task: popularity prediction. Besides, the superior of the coupled structure in CoupledGNN over single-GNN is also demonstrated. Finally, the effect of hyper-parameters or experimental settings is analyzed comprehensively.

\subsection{Overall Performance}
The experimental results for popularity prediction on both the synthetic data and real-world data in Sina Weibo are shown in Table~\ref{tab:popularity-synthetic} and Table~\ref{tab:popularity-sina} respectively. 

For SEISMIC, due to it predicts infinite popularity for some pieces of information, we only use mRSE and WroPerc as the evaluation metrics for a fair comparison. From the experimental results, we can see that SEISMIC performs not well on both data sets. Since it only estimates the impact of the cascading effect in each generation by the average number of fans, it's easy to deviate from complex and real situations, thus having limited predictive power. As for DeepCas, the deep representation learning methods, it does perform better than the feature-based methods. This result indicates that it's effective to automatically learning the representation of the cascade graph through an end-to-end manner rather than heuristically design hand-crafted features with prior knowledge.

As for our CoupledGNN model, it outperforms all the baselines on both synthetic and real-world datasets, achieving more than 10\% improvement over DeepCas in Sina Weibo under the MRSE. These results demonstrate that it's effective to utilize graph neural networks to capture the cascading effect along network structure and to predict the popularity of online content on social platforms. In other words, considering the interactions between early adopters and the potential active users, as well as the interactions among potential active users over network structure is useful to further improve the prediction performance for future popularity.

As for the observation time in Sina Weibo (Figure~\ref{tab:popularity-sina}), we can see that the longer the observation time is, the smaller the errors are (MRSE, MAPE, WrongPerc). This is applicable to all the methods. The reason is that the longer the observation time is, the more information is available, making the prediction easier.

\begin{table}
  \caption{Compare CoupledGNN with Single-GNN}
  \label{tab:singlegnn}
  \begin{tabular}{c|c|c|c}
    \toprule
    Observation Time & \multicolumn{3}{c}{1 hour}\\
     \midrule
     Evaluation Metric & MRSE &MAPE & WroPerc   \\
    \midrule
         Single-GCN &0.1964 &0.3707 &29.11\%\\
         Single-GAT &0.1999 &0.3754 &30.82\% \\
         CoupledGNN 
                    &\textbf{0.1816} &\textbf{0.3515} &\textbf{25.68\%} \\
    \midrule
    Observation Time & \multicolumn{3}{c}{2 hours}\\
     \midrule
     Evaluation Metric & MRSE &MAPE & WroPerc   \\
    \midrule
         Single-GCN &0.1595 &0.3201 &22.26\%\\
         Single-GAT &0.1569 &0.3199 &20.55\%\\
         CoupledGNN 
                    &\textbf{0.1397} &\textbf{0.2989} &\textbf{17.81\%}\\
        \midrule
    Observation Time & \multicolumn{3}{c}{3 hours}\\
     \midrule
     Evaluation Metric & MRSE &MAPE & WroPerc   \\
    \midrule
         Single-GCN &0.1230 &0.2653 &16.10\%\\
         Single-GAT &0.1222 &0.2655 &16.10\% \\
         CoupledGNN 
                    &\textbf{0.1120} &\textbf{0.2611} &\textbf{13.01\%} \\
  \bottomrule
\end{tabular}
\end{table}

\subsection{Compare CoupledGNN with Single-GNN}
To further demonstrate the advantages of our CoupledGNN structure, we simplify our method with two versions: Single-GCN and Single-GAT. In both these two simplified versions, we concatenate the activation state and influence representation as one vector associated with each user. Then a single graph neural network, i.e., the commonly used graph convolution network (GCN)~\cite{kipf2017semi} or graph attention neural networks (GAT)~\cite{velivckovic2018graph} are applied to iteratively update the vector associated with each user. The final popularity is obtained similarly as CoupledGNN, i.e., applying a sum pooling mechanism over all users after transforming the vector of each user at the last layer into one-dimensional value. Other hyper-parameters and the implementation details are the same as CoupledGNN. 

The experimental results are shown in Table~\ref{tab:singlegnn}. 
Single-GCN and Single-GAT perform almost similarly, indicating that when modeling the future popularity, the normalized Laplacian matrix used in GCN is already a good reflection of the correlation between pair of users with a linking edge. The attention mechanism adopted in GAT won't significantly improve the performance further. 
As for our CoupledGNN, it significantly improves the prediction performance under all the evaluation metrics compared with the Single-GNN methods. These results demonstrate that the activation state and influence representation play different roles in the modeling of future popularity. Instead of mixing them up together, it's effective to model the activation state and influence representation by two graph neural networks respectively and then couple them by gating mechanisms.


\subsection{Parameter Analysis}
We further analyze the effect of the coefficient $\lambda$, the number of layers in CoupledGNN in this subsection. The influence of a partial lack of network is also analyzed.

\subsubsection{The coefficient $\lambda$ in loss function}
We vary the coefficient $\lambda$ from $0.0,0.5,1.0,10.0$ to $20.0$, and the corresponding mean relative square loss is $0.1109, 0.1101, 0.1111,0.1111,0.1141$. In other words, the MRSE loss is first reduced with the increasing of $\lambda$, indicating that adding the user-level cross entropy loss is beneficial for macro popularity prediction. 
However, with the continuous increase of $\lambda$, the model pays too much attention to the user-level prediction while ignoring the macro popularity prediction task, thus resulting in less effective prediction performance. 


\begin{figure}
  \centering
  \includegraphics[width=1.65in]{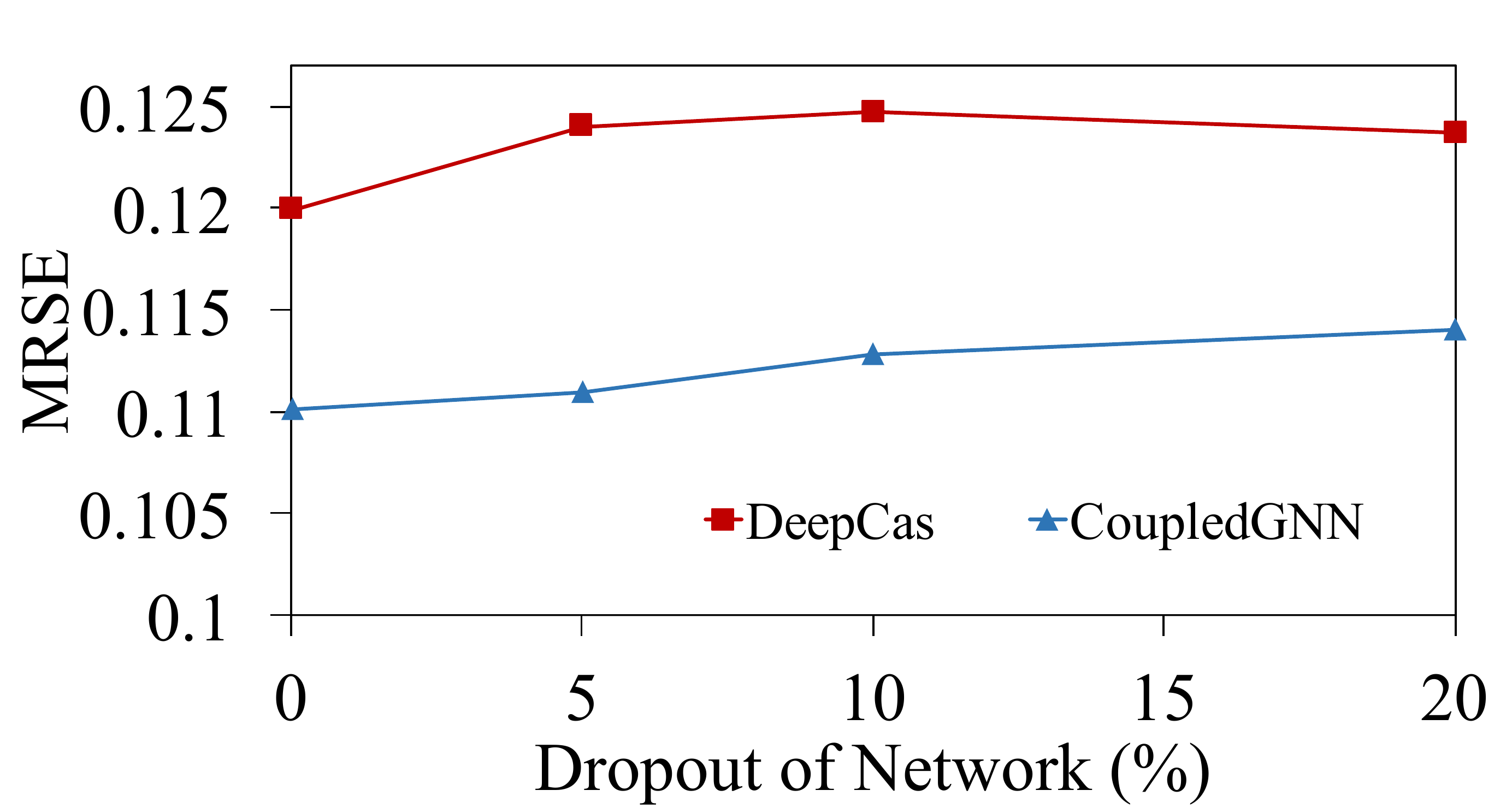}
  \includegraphics[width=1.65in]{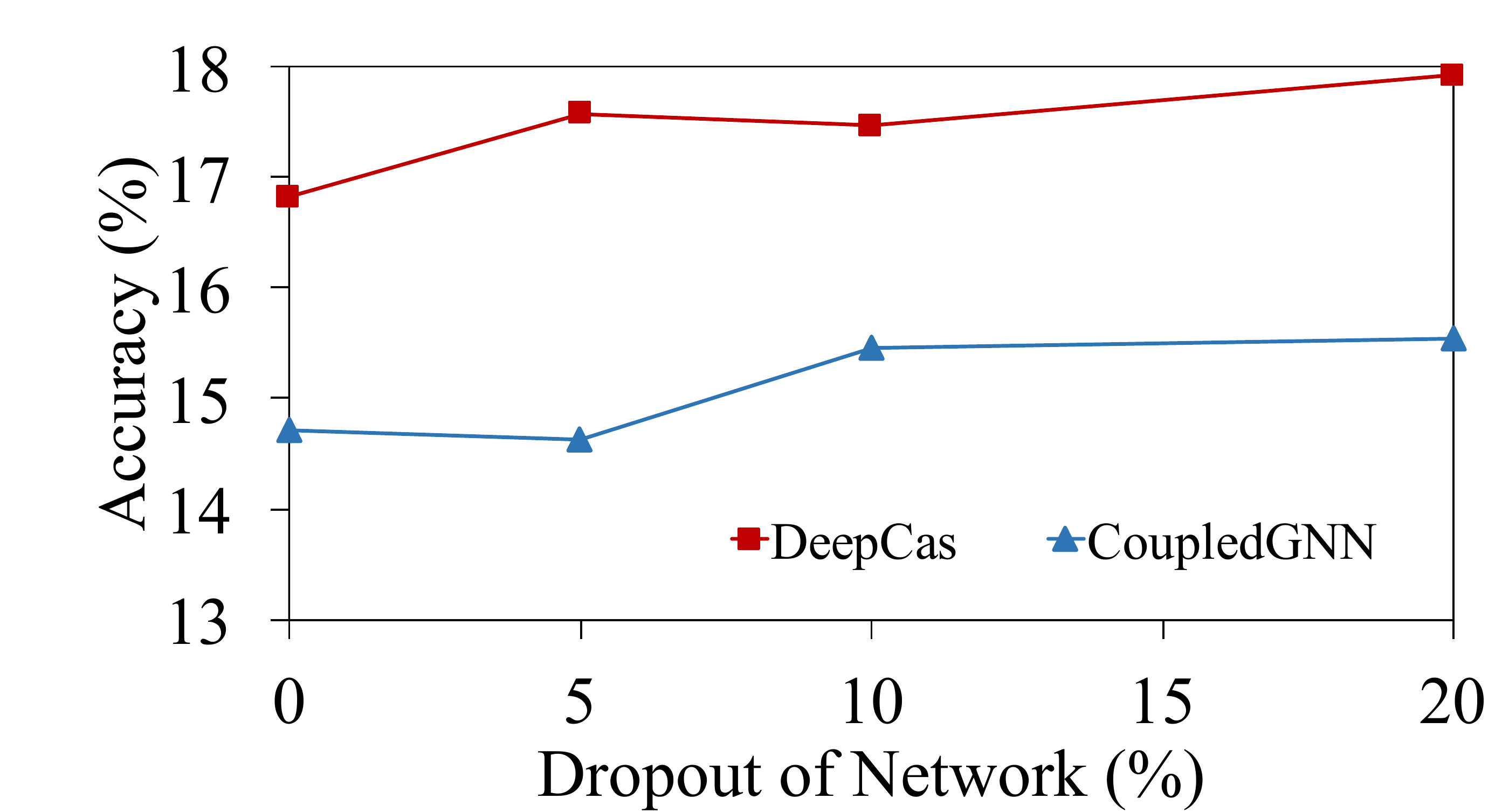}
  \caption{\label{dropoutofnetwork}The influence of a partial lack of network.}
\end{figure}

\subsubsection{The influence of a partial lack of network}
Considering that our method is based on the given underlying network, we further construct an experiment with the partial lack of network to better demonstrate the applicability and generality of our methods. Specifically, we randomly dropout a certain percentage of edges in the social network under the premise of network connectivity. Then we train and test the prediction model based on such an incomplete network. Note that, not only our CoupledGNN are influenced by such dropout of network edges, but also the baseline methods. Here, we compare our methods with the strong baseline, i.e., DeepCas, while varying the dropout of the network from 0\%, 5\%,10\% to 20\%. Figure~\ref{dropoutofnetwork} shows that both the performance of our CoupledGNN and DeepCas will be slightly degraded by the dropout of the network. But by comparison, our methods always significantly perform better than DeepCas. In conclusion, our CoupledGNN model is still suitable for predicting the future popularity of online content even when a small part of the network structure is lacking.

\begin{figure}
  \centering
  \includegraphics[width=2.1in]{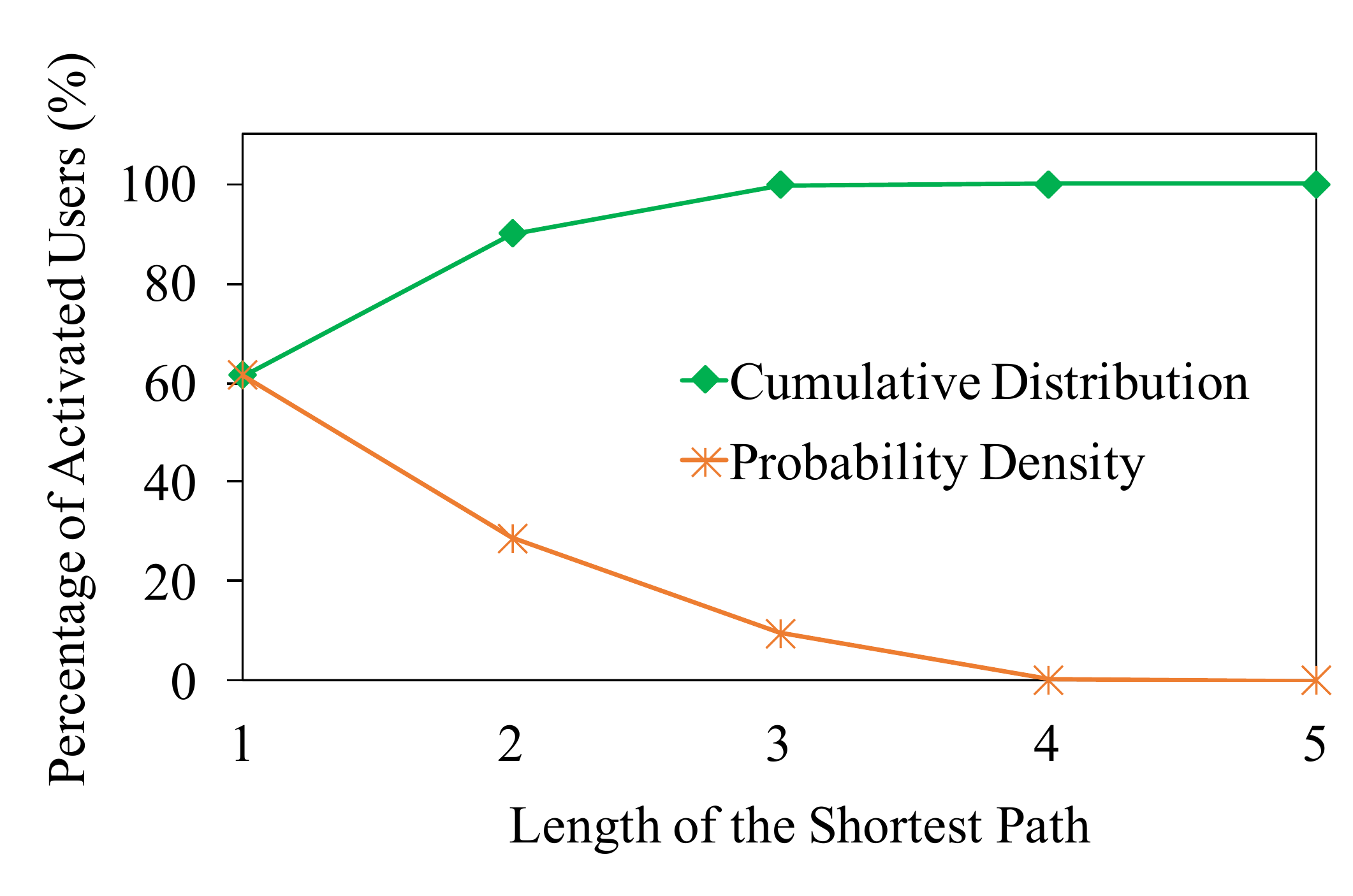}
  \caption{\label{shortestpath} The distribution of the shortest path length between activated users after the observation time and the set of early adopters within observation time.}
\end{figure}

\subsubsection{The number of layers in CoupledGNN}
While we apply our CoupledGNN to capture the cascading effect along the network, it's interesting to further find out whether the number of layers $K$ in our CoupledGNN can correspond to the scope of the cascading effect. 
We take the Sina Weibo as a showing case. On the one hand, as mentioned in Section 5.2, the number of layers $K$ is chosen from ${2,3,4}$. When the observation time window is 1 hour, we obtain the best performance on the validation set with $K =3$. On the other hand, Figure~\ref{shortestpath} shows the distribution of the length of the shortest path from the set of \textbf{early adopters} within observation time to the \textbf{activated users} after the observation time. Such distribution reflects the range of cascading effect caused by the early adopters. We can see that almost all the activated users, i.e., 99.76\%, can be covered within three-hops in the neighborhood of early adopters. This means that the optimal number of layers obtained by our methods can exactly match the scope of the cascading effect in this case, which provides guidance for setting the hyper-parameter $K$ in other situations, as well as further supports the effectiveness of our proposed method.

\section{Conclusion}
In this paper, we focus on the problem of network-aware popularity prediction of online content on social platforms. How to capture the cascading effect is one of the keys to accurately predict future popularity and tackle this problem. Inspired by the success of graph neural networks on various non-Euclidean domains, we propose CoupledGNN to characterize the critical cascading effect along the network structure. We devote to modeling the two crucial components in the cascading effect, i.e., the iterative interplay between node activation states and the spread of influence, by two coupled graph neural networks respectively. Specifically, one graph neural network models the interpersonal influence, gated by the activation state of users. The other graph neural network models the activation state of users via interpersonal influence from their neighbors. 
The iterative update mechanism of neighborhood aggregation in GNNs effectively captures such a cascading effect in popularity prediction along the underlying network. The experiments conducted on both synthetic and real-world data validate the effectiveness of our proposed method for popularity prediction.
As for future work, we will devote to modeling the cascading effect along the network when further given the specific adoption time of early adopters.

%
\begin{acks}
This work is funded by the National Natural Science Foundation of China under grant numbers 61425016, 61433014, 91746301, and 61472400. This work is supported by Beijing Academy of Artificial Intelligence (BAAI). Huawei Shen is also funded by K.C. Wong Education Foundation and the Youth Innovation Promotion Association of the Chinese Academy of Sciences.
\end{acks}

%

%

\end{document}